\documentclass[aps,twocolumn,pra,superscriptaddress,showpacs,tightenlines]{revtex4-1}
\usepackage{amsmath}
\usepackage{graphicx}
\usepackage{epsfig}
\usepackage{amsfonts}
\usepackage{txfonts}

\begin{document}
\title{Single-photon quadratic optomechanics}
\author{Jie-Qiao Liao}
\affiliation{CEMS, RIKEN, Saitama 351-0198, Japan}
\author{Franco Nori}
\affiliation{CEMS, RIKEN, Saitama 351-0198, Japan}
\affiliation{Physics Department, The University of Michigan, Ann Arbor, Michigan 48109-1040, USA}
\date{\today}
\begin{abstract}
We present exact analytical solutions to study the coherent interaction
between a single photon and the mechanical motion of a membrane in quadratic optomechanics.
We consider single-photon emission and scattering when the photon is initially inside the cavity and in the fields outside the cavity, respectively.
Using our solutions, we calculate the single-photon emission and scattering spectra,
and find relations between the spectral features and the system's inherent parameters,
such as: the optomechanical coupling strength, the mechanical frequency, and the cavity-field decay rate.
In particular, we clarify the conditions for the phonon sidebands to be visible.
We also study the photon-phonon entanglement for the long-time emission and scattering states. The linear entropy is employed to characterize
this entanglement by treating it as a bipartite one between a single mode of phonons and a single photon.
\end{abstract}

\maketitle

\section{Introduction}

The hybrid coherent coupling\cite{Xiang2013} between electromagnetic and mechanical degrees of freedom
is at the heart of cavity optomechanics\cite{Kippenberg2008,Marquardt2009,Aspelmeyer2013}. In general,
optomechanical couplings can be classified into two categories: linear or quadratic couplings.
Namely, the coupling term is proportional to either $x$ or $x^{2}$ ($x$ being
the mechanical displacement). For a mechanical resonator, a linear coupling corresponds to a force acting upon the mechanical resonator and this leads to a displacement of its equilibrium
position. However, for a quadratic coupling, it will change the resonant frequency of the mechanical resonator (in the new representation, rather than in the original representation).
This is because the quadratic-coupling term can be integrated into the potential energy of the harmonic oscillator (changing the effective stiffness of a spring), and hence
the frequency of the mechanical resonator is renormalized. This renormalized harmonic oscillator is related to the original one by a squeezing transformation.

To better understand and exploit optomechanical couplings, it is highly desirable to realize these
couplings in the single-photon strong-coupling regime, in which the couplings involving a single
photon can produce observable effects on both mechanical and electromagnetic signals. Such a
regime is important to test the fundamentals of quantum theory\cite{Law1995,Mancini1997,Bose1997,Bouwmeester2003}
and to explore possible applications of optomechanical devices to future quantum technology\cite{Stannigel2011,Stannigel2012,Ludwig2012}.
In the past several years, much attention has been paid to the single-photon strong-coupling
regime of linear coupling\cite{Gupta2007,Murch2008,Brennecke2008,Eichenfield2009,Rabl2011,Nunnenkamp2011,Chenp2011,Liao2012,He2012,Liu2013,Marquardt2013,Liao2013,Xu2013,Lu2012}.
Considerable theoretical studies, such as photon statistics\cite{Rabl2011,Nunnenkamp2011,Liu2013,Marquardt2013,Liao2013} and mechanical-state engineering\cite{Xu2013}, have been carried out in this regime.
Two theoretical proposals\cite{Sillanpaa2013,Rimberg2014} have recently been suggested to reach this regime using superconducting circuits with Josephson junctions.
In addition, experimental advances in linear optomechanics are being made towards the single-photon strong-coupling regime\cite{Gupta2007,Murch2008,Brennecke2008,Eichenfield2009}.
However, for the quadratic coupling, though much attention has been paid to this area\cite{Harris2008,Meystre2008,Meystre2008B,Agarwal2008,Jayich2008,Harris2010,Nunnenkamp2010,Stamper-Kurn2010,Vanner2011,Agarwal2011,Biancofiore2011,Cheung2011,Deng2012,Xiao2013,Flowers-Jacobs2012,Meystre2012,Paternostro2013,Bhattacharya2013,Liao2013B,Tan2013,Yang2013},
not much work has been devoted to the single-photon strong-coupling regime because the currently attainable coupling strength is weak.
Recently, some methods have been proposed to increase the quadratic optomechanical coupling strength or to seek other possible realization of quadratic optomechanics.
For example, an experiment\cite{Flowers-Jacobs2012} demonstrated that the quadratic coupling strength can be
increased significantly using a fiber cavity with a smaller mode size, and a smaller and lighter membrane.
A measurement-based method has also been proposed to obtain an effective quadratic optomechanics\cite{Vanner2011}.
In addition, some other systems, such as trapped cold atoms or a dielectric nano- or
microparticle, have been suggested to simulate an effective quadratic optomechanical coupling\cite{Paternostro2013}.
These works provide a possibility of studying the quantum nonlinearity in quadratic optomechanics.
Motivated by these advances, it is of interest to study the quadratic optomechanical coupling in the
single-photon strong-coupling regime.

When a quadratic optomechanical cavity works in the single-photon strong-coupling
regime, the frequency change of the mechanical resonator
induced by a single photon will, in turn, significantly affect the cavity field, causing some observable
features in the cavity photon spectrum. Thus, a natural question arises: how the spectrum may
characterize the single-photon strong-coupling regime? In this paper, we answer this question
by calculating analytically the spectrum of single-photon emission and scattering.
In particular, we build a connection between
the spectral features and the system's inherent parameters. We also clarify the condition for
observing the phonon sidebands in the spectra. It should be pointed out that below we
assume that the mechanical resonator has been pre-cooled to a low-phonon-number regime.
Moreover, in typical optomechanical systems, the optical decay rate $\gamma_{c}$ is much larger than the mechanical decay rate $\gamma_{M}$.
Under these two conditions, we will only consider the optical dissipation
and neglect the mechanical dissipation.
This is because the emission and scattering processes will be
completed in a time interval $1/\gamma_{c}\ll t\ll 1/\gamma_{M}$. During this period the mechanical dissipation is negligible.
In this issue, the emission and pulse-scattering cases are different from the continuous-wave driving case.
When the system is driven by a continuous-wave field, the mechanical dissipation should be included for the steady-state solution\cite{Liao2013B}.

Accompanying the processes of single-photon absorbtion and emission, the total system experiences transitions
involving phonon sidebands, and hence the frequency of the emitted photon will be related to the states of the phonon sidebands
due to energy conservation. This relation leads to the generation of photon-phonon entanglement.
Since the emitted photon exists in the continuous modes of the outside fields, this entangled state
involves a single mode of phonons and a set of continuous modes of a single photon. In general, it is hard
to characterize such type of entanglement. However, from the point of view of a single photon rather than photon modes,
we could treat this entanglement as a bipartite one between a single photon and the phonon mode. For a pure initial state,
the linear entropy can be employed to describe this entanglement.

\section{The model}

\begin{figure}[tbp]
\center
\includegraphics[bb=29 404 302 763, width=3.3 in]{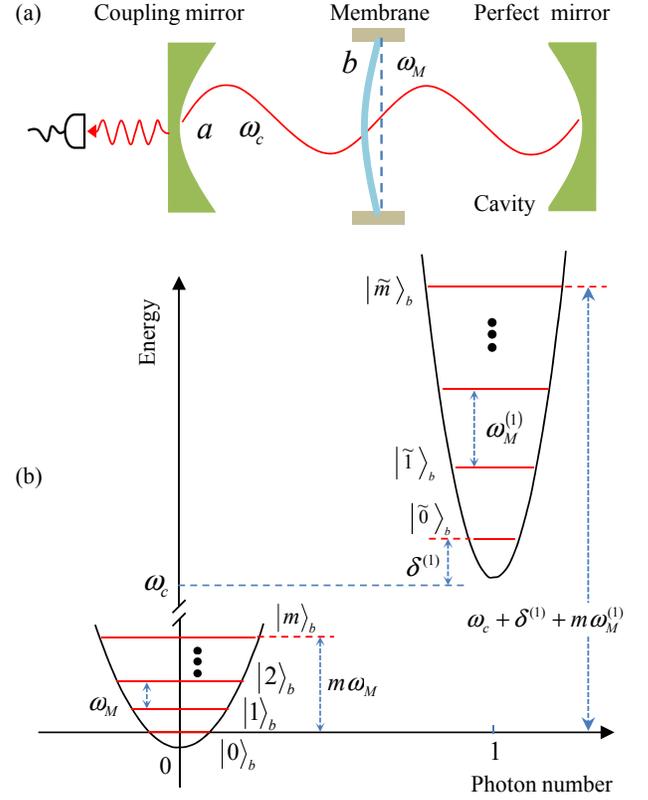}
\caption{(Color online) (a) Schematic diagram of a quadratic optomechanical system
with a ``membrane-in-the-middle" configuration.
(b) The diagram of the energy-level structure (unscaled) of the optomechanical system
when the cavity is in a vacuum or contains a single photon.}
\label{setup}
\end{figure}

We consider a quadratic optomechanical system with a
``membrane-in-the-middle" configuration [see Fig.~\ref{setup}(a)], where a thin dielectric
membrane is placed inside a Fabry-P\'{e}rot cavity. We model the moving membrane as a
harmonic oscillator and focus on a single-mode field in the cavity. When the membrane is
placed at a node (or antinode) of the intracavity standing wave, the cavity field will quadratically couple to the mechanical motion of the membrane.
Let us denote the position and momentum operators of the membrane as $x$ and $p$, then the Hamiltonian of the system is (with $\hbar=1$)
\begin{equation}
H_{\rm opc}=\omega_{c}(x)(a^{\dagger}a+1/2)+\frac{p^{2}}{2M}+\frac{1}{2}M\Omega^{2}x^{2},\label{Hameq1}
\end{equation}
where $a^{\dagger}$ and $a$ are the creation and annihilation operators of the single-mode cavity field, respectively.
In the quadratic coupling case, the cavity-field frequency depends on the mechanical motion by
$\omega_{c}(x)=\omega_{c}+\eta x^{2}$, with $\eta=\frac{1}{2}\frac{\partial^{2}\omega_{c}(x)}{\partial x^{2}}|_{x=0}$, where $\omega_{c}$
is the cavity-field frequency when the membrane is at rest. The parameters $M$ and $\Omega$ in Eq.~(\ref{Hameq1}) are the mass and frequency of the mechanical mode.
By reorganizing the coupling term between the zero-point energy of the optical mode and the mechanical motion into the mechanical potential energy, we have
\begin{equation}
H_{\rm opc}=\omega_{c}a^{\dagger }a+\frac{p^{2}}{2M}+\frac{1}{2}M\omega
_{M}^{2}x^{2}+\eta a^{\dagger }ax^{2}+\frac{1}{2}\omega _{c},\label{Hameq2}
\end{equation}
where $\omega_{M}=\sqrt{\Omega^{2}+\eta/M}$ is the renormalized frequency of the membrane. We note that it is justified to work
in the representation of the mechanical frequency $\omega_{M}$, because the coupling term $\eta x^{2}/2$ always exists.
By introducing the mechanical creation and annihilation operators $b^{\dagger}$ and $b$, by
$x=\sqrt{1/(2M\omega_{M})}(b^{\dagger }+b)$ and $p=i\sqrt{M\omega_{M}/2}(b^{\dagger}-b)$,
the Hamiltonian, up to a constant term $(\omega_{c}+\omega_{M})/2$, becomes\cite{Harris2008}
\begin{equation}
H_{\rm opc}=\omega_{c}a^{\dagger}a+\omega_{M}b^{\dagger}b+g_{0}a^{\dagger}a(b^{\dagger}+b)^{2},\label{Hameq3}
\end{equation}
where $g_{0}=\eta/(2M\omega _{M})$. The third term in Eq.~(\ref{Hameq3}) describes a
quadratic optomechanical coupling with a strength $g_{0}$ between the cavity field and the membrane.
We point out that since the reorganized frequency $\omega_{M}$ of the mechanical mode also depends on the coupling strength $g_{0}$
by $\omega _{M}=\sqrt{\Omega^{2}+2g_{0}\omega_{M}}$, the ratio of the coupling strength $g_{0}$ over the mechanical frequency $\omega_{M}$
is bounded by $g_{0}/\omega_{M}<1/2$. However, below we also consider parameters by extending the scope of the ratio beyond this bound. This is because the bound can be exceeded
in some other quadratical optomechanical systems, such as trapped cold atoms.

When there are $s$ photons in the cavity, the last two terms in Hamiltonian~(\ref{Hameq3}) can be
renormalized as a harmonic-oscillator Hamiltonian with the resonant
frequency $\sqrt{\omega_{M}(\omega_{M}+4sg_{0})}$. To keep the stability of the membrane (i.e., the frequency should be a positive number),
the strength $g_{0}$ should satisfy the condition $(\omega_{M}+4sg_{0})>0$.
The photon number operator $a^{\dagger}a$ in Hamiltonian $H_{\rm opc}$ is a conserved quantity,
and hence for a given photon number $s$, the coupling actually takes a quadratic form
$sg_{0}(b^{\dagger}+b)^{2}$, which can be diagonalized with the single-mode squeezing transformation. Denoting the
harmonic-oscillator number states of the cavity field and the membrane as $|s\rangle_{a}$ and $|m\rangle_{b}$
($s,m=0,1,2,...$) respectively, then the eigensystem of the Hamiltonian $H_{\rm opc}$ can be obtained as
\begin{equation}
H_{\rm opc}\vert s\rangle_{a}\vert\tilde{m}(s)\rangle_{b}=(s\omega_{c}+\delta^{(s)}+m\omega_{M}^{(s)})\vert s\rangle_{a}\vert\tilde{m}(s)\rangle_{b},\label{Eigeq4}
\end{equation}
where we introduce the $s$-photon coupled membrane's resonant frequency $\omega_{M}^{(s)}$ and energy-level shift $\delta^{(s)}$,
\begin{equation}
\omega_{M}^{(s)}=\omega_{M}\sqrt{1+\frac{4sg_{0}}{\omega_{M}}},\hspace{0.5 cm} \delta^{(s)}=\frac{1}{2}(\omega_{M}^{(s)}-\omega_{M}).\label{Eigfreqeq5}
\end{equation}
The $s$-photon squeezed phonon number state in Eq.~(\ref{Eigeq4}) is defined by
\begin{equation}
\vert\tilde{m}(s)\rangle_{b}=S_{b}(\eta^{(s)})\vert m\rangle_{b},\label{squnumstaeq6}
\end{equation}
where $S_{b}(\eta^{(s)})=\exp[\frac{\eta^{(s)}}{2}(b^{2}-b^{\dagger2})]$ is a squeezing operator with the squeezing factor
\begin{equation}
\eta^{(s)}=\frac{1}{4}\ln\left(1+\frac{4sg_{0}}{\omega_{M}}\right).\label{etaseq7}
\end{equation}
We note that the eigensystem of the Hamiltonian~(\ref{Hameq3}) has been derived in previous studies\cite{Agarwal2008,Bhattacharya2013,Liao2013B}.
In the zero-photon case, we have $\vert\tilde{m}(0)\rangle_{b}=\vert m\rangle_{b}$, $\omega_{M}^{(0)}=\omega_{M}$, and $\delta^{(0)}=0$.
For following convenience, the energy-level structure of the system in the zero- and one-photon cases is shown in Fig.~\ref{setup}(b).

To include the dissipation of the cavity field, we assume that the cavity photons can couple with the outside
fields through the coupling mirror. Without loss of generality, we model the environment of the cavity field as
a harmonic-oscillator bath. Then the Hamiltonian of the whole system including the optomechanical cavity and
the environment can be written as
\begin{equation}
H=H_{\rm opc}+\int_{0}^{\infty}\omega_{k}c_{k}^{\dagger}c_{k}dk+\xi\int_{0}^{\infty}(ac_{k}^{\dagger}+c_{k}a^{\dagger})dk,\label{Htoteq8}
\end{equation}
where the annihilation operator $c_{k}$ describes the $k$th mode of the
outside fields with resonant frequency $\omega_{k}$. The coupling
between the cavity field and the outside fields is described by the photon-hopping
interaction with strength $\xi$. Since the decay rate $\gamma_{M}$ of the mechanical resonator is much smaller than the decay rate $\gamma_{c}$ of the
cavity, then during the emission and wave-packet scattering time interval $1/\gamma_{c}\ll t\ll1/\gamma_{M}$, the damping of the membrane is negligible.
In this work we take into account the dissipation of the cavity and neglect the mechanical dissipation.

\section{States in the single-photon subspace}

In the rotating frame with respect to
$H_{0}=\omega_{c}a^{\dagger}a+\omega_{c}\int_{0}^{\infty}c_{k}^{\dagger}c_{k}dk$,
the Hamiltonian~(\ref{Htoteq8}) becomes
\begin{eqnarray}
H_{I}&=&\omega_{M}b^{\dagger}b+g_{0}a^{\dagger}a(b^{\dagger}+b)^{2}
+\int_{0}^{\infty}\Delta_{k}c_{k}^{\dagger}c_{k}dk\nonumber\\
&&+\xi\int_{0}^{\infty}(ac_{k}^{\dagger}+c_{k}a^{\dagger})dk,\label{Htotroteq9}
\end{eqnarray}
where $\Delta_{k}=\omega_{k}-\omega_{c}$ is the detuning of the $k$th mode photon from the cavity frequency.
The total photon number, $N=a^{\dagger}a+\int_{0}^{\infty}c_{k}^{\dagger }c_{k}dk$,
in the whole system is a conserved quantity because of
$[N,H_{I}]=0$. Denoting $\vert\tilde{m}(1)\rangle_{b}\equiv\vert\tilde{m}\rangle_{b}$ for conciseness,
a general state in the single-photon subspace of the total system can be
written as
\begin{eqnarray}
\vert\varphi(t)\rangle&=&\sum_{m=0}^{\infty}A_{m}(t)\vert1\rangle_{a}\vert\tilde{m}
\rangle_{b}\vert\emptyset\rangle\nonumber\\
&&+\sum_{m=0}^{\infty}\int_{0}^{\infty}B_{m,k}(t)\vert0\rangle_{a}\vert m\rangle_{b}\vert 1_{k}\rangle dk,\label{singphostaeq10}
\end{eqnarray}
where $\vert 1\rangle_{a}\vert\tilde{m}\rangle_{b}\vert\emptyset\rangle$ denotes the state
with the cavity in the single-photon state $\vert 1\rangle_{a}$, the membrane in the single-photon squeezed number state
$|\tilde{m}\rangle_{b}$ (hereafter we call it as squeezed number state for conciseness), and the outside fields in a vacuum $\vert\emptyset\rangle$.
Also $\vert 0\rangle_{a}\vert m\rangle_{b}\vert
1_{k}\rangle$ denotes the state with a vacuum cavity field $\vert 0\rangle_{a}$, the membrane
in the number state $|m\rangle_{b}$, and one
photon in the $k$th mode of the outside fields $\vert 1_{k}\rangle$. The variables $A_{m}(t)$ and $B_{m,k}(t)$ are probability amplitudes.

We point out that these squeezed number states in Eq.~(\ref{singphostaeq10}) satisfy
the completeness $\sum_{m=0}^{\infty}|\tilde{m}\rangle_{b}\,_{b}\langle\tilde{m}|=I_{b}$ ($I_{b}$ is the
identity operator in the Hilbert space of mode $b$) and orthogonality $_{b}\langle\tilde{m}|\tilde{n}\rangle_{b}=\delta_{m,n}$.
Moreover, the overlap $_{b}\langle m\vert\tilde{n}\rangle_{b}=\,_{b}\langle m\vert S_{b}(\eta^{(1)})\vert
n\rangle_{b}$ between the squeezed number state $|\tilde{n}\rangle_{b}$ and the harmonic-oscillator number state $|m\rangle_{b}$
is determined by the relation
\begin{eqnarray}
_{b}\langle m\vert S_{b}(\eta^{(1)})\vert n\rangle_{b}
&=&\frac{\sqrt{m!n!}}{(\cosh\eta^{(1)})^{n+1/2}}
\sum\limits_{l^{\prime}=0}^{\textrm{Floor}[\frac{m}{2}
]}\sum\limits_{l=0}^{\textrm{Floor}[\frac{n}{2}]}\frac{(-1)^{l^{\prime}}}{l!l^{\prime}!}\nonumber\\
&&\times\frac{(\frac{1}{2}\tanh\eta^{(1)})^{l+l^{\prime}}}{(n-2l)!}(\cosh\eta^{(1)})^{2l}\delta_{m-2l^{\prime},n-2l},\label{overlapeq11}\nonumber\\
\end{eqnarray}
where $\eta^{(1)}=(1/4)\ln(1+4g_{0}/\omega_{M})$ and the function Floor$[x]$ gives the greatest integer less than or equal to $x$.
In principle, the state of the whole system can be obtained by solving the Schr\"{o}dinger equation under
a given initial condition. Below we will consider single-photon emission and scattering.

According to Eqs.~(\ref{Htotroteq9}), (\ref{singphostaeq10}), and the Schr\"{o}dinger equation	
$i\frac{\partial}{\partial t}\vert\varphi(t)\rangle=H_{I}\vert\varphi(t)\rangle$,
we obtain the equations of motion for the probability amplitudes
\begin{subequations}
\label{eqofmotioneq12}
\begin{align}
\dot{A}_{m}(t)=&-i(\delta^{(1)}+m\omega_{M}^{(1)})A_{m}(t)\nonumber\\
&-i\xi\sum_{n=0}^{\infty}\int_{0}^{\infty}\,_{b}\langle\tilde{m}\vert n\rangle_{b}\,B_{n,k}(t)dk,\\
\dot{B}_{m,k}(t)=&-i(\Delta_{k}+m\omega_{M})B_{m,k}(t)\nonumber\\
&-i\xi\sum_{n=0}^{\infty}\,_{b}\langle m\vert \tilde{n}\rangle_{b}\,A_{n}(t),
\end{align}
\end{subequations}
where the single-photon coupled membrane's frequency and energy-level shift are given by $\omega_{M}^{(1)}=\omega_{M}\sqrt{1+4g_{0}/\omega_{M}}$
and $\delta^{(1)}=(\omega_{M}^{(1)}-\omega_{M})/2$. In addition, the coefficients $_{b}\langle\tilde{m}\vert n\rangle_{b}$
and $_{b}\langle m\vert \tilde{n}\rangle_{b}$ can be calculated using Eq.~(\ref{overlapeq11}).

The equations of motion (\ref{eqofmotioneq12}) for the probability amplitudes may be solved with the Laplace transform method under
a given initial condition. In the single-photon emission case, a single photon is initially inside the cavity, and the outside fields are in a vacuum.
For the mechanical mode, we first assume that its initial state is an arbitrary number state $\vert n_{0}\rangle_{b}$.
Based on the solution in this case, the solution for the general initial membrane state can be obtained accordingly by superposition.
For the initial state $\vert 1\rangle_{a}\vert n_{0}\rangle_{b}\vert\emptyset \rangle$, the corresponding initial condition for
Eq.~(\ref{eqofmotioneq12}) is $A_{m}(0)=\,_{b}\langle\tilde{m} \vert n_{0}\rangle_{b}$ and $B_{m,k}(0)=0$. In the single-photon scattering case, the single photon is initially in a
Lorentzian wave packet in the outside fields, and the cavity is in a vacuum $\vert0\rangle_{a}$.
We also assume that the membrane is initially in the number state $\vert n_{0}\rangle_{b}$, and then
the initial condition for Eq. (\ref{eqofmotioneq12})  becomes
$A_{m}(0)=0$ and $B_{m,k}(0)=\sqrt{\frac{\epsilon}{\pi}}\frac{1}{\Delta_{k}-\Delta_{0}+i\epsilon}\delta_{m,n_{0}}$,
where $\Delta_{0}$ and $\epsilon$ are the detuning center and spectral width of the photon, respectively.
Below, we will consider the single-photon emission and scattering, respectively.

\section{Single-photon emission}

In the single-photon emission case, a single photon is initially inside the cavity, and the outside fields are in a vacuum.
Without loss of generality, we assume that the initial state of the membrane is an arbitrary number state $\vert n_{0}\rangle_{b}$. Once the solution in
this case is obtained, the solution for the general initial membrane state can be obtained accordingly by superposition.
In this case, with the Laplace transform method, we obtain the long-time solution for these probability amplitudes as $A_{n_{0},m}(\infty)=0$ and
\begin{eqnarray}
B_{n_{0},m,k}(\infty)=\sum_{n=0}^{\infty}\frac{\sqrt{\gamma_{c}/(2\pi)}\;\,_{b}\!\langle m\vert \tilde{n}\rangle_{b}\,_{b}\langle\tilde{n}\vert n_{0}\rangle_{b}\,
e^{-i(\Delta_{k}+m\omega_{M})t}}{\Delta_{k}-\delta^{(1)}-n\omega_{M}^{(1)
}+m\omega_{M}+i\gamma_{c}/2},\label{emislontsoleq13}
\end{eqnarray}
where we introduce the cavity-field decay rate $\gamma_{c}=2\pi\xi^{2}$ and add the subscript $n_{0}$ in $A_{n_{0},m}(t)$ and $B_{n_{0},m,k}(t)$ to mark the membrane's
initial state $|n_{0}\rangle_{b}$.

In the long-time limit, the single photon completely leaks out of the cavity and hence the cavity is in
a vacuum [$A_{n_{0},m}(\infty)=0$]. The amplitude $B_{n_{0},m,k}(\infty)$ exhibits a clear physical picture
for the single-photon emission. Specifically, the initial state $|1\rangle_{a}|n_{0}\rangle_{b}$ can be
expanded as $\sum_{n=0}^{\infty}c_{n_{0},n}
|1\rangle_{a}\vert\tilde{n}\rangle_{b}$, with $c_{n_{0},n}=\,_{b}\langle\tilde{n} \vert n_{0}\rangle_{b}$.
For each component $|1\rangle_{a}\vert\tilde{n}\rangle_{b}$, the single-photon emission
process induces the transition $|1\rangle_{a}\vert\tilde{n}\rangle_{b}\rightarrow|0\rangle_{a}
\vert m\rangle_{b}$. The corresponding
transition amplitude is proportional to the numerator in Eq.~(\ref{emislontsoleq13}).
Due to the quadratic terms of $b$ and $b^{\dagger}$ in $S_{b}(\eta^{(1)})$, $|m\rangle_{b}$ and $\vert\tilde{n}\rangle_{b}$
should have the same parity, i.e., being odd or even. Consequently, the phonon number distribution in the long-time
state of the membrane will have the same parity as its initial component $\vert n_{0}\rangle_{b}$. In addition,
we can derive the resonant condition in this emission process from the energy-level structure in Fig.~\ref{setup}(b).
For the transition $|1\rangle_{a}\vert\tilde{n}\rangle_{b}\rightarrow|0\rangle_{a}\vert m\rangle_{b}$,
the frequency of the emitted photon is $\omega_{k}=\omega_{c}+\delta^{(1)}+n\omega_{M}^{(1)}-m\omega_{M}$,
which is consistent with the resonance condition
\begin{eqnarray}
\Delta_{k}=\delta^{(1)}+n\omega_{M}^{(1)}-m\omega_{M},\label{emisrescondeq14}
\end{eqnarray}
obtained from the pole of the denominator in Eq.~(\ref{emislontsoleq13}).

We know from Eqs.~(\ref{singphostaeq10}) and (\ref{emislontsoleq13}) that, corresponding to the initial state
$|1\rangle_{a}|n_{0}\rangle_{b}|\emptyset\rangle$, the long-time state of the whole system is
\begin{eqnarray}
\vert\varphi_{n_{0}}(\infty)\rangle
=\sum_{m=0}^{\infty}\int_{0}^{\infty}B_{n_{0},m,k}(\infty)\vert
0\rangle_{a}\vert m\rangle_{b}\vert1_{k}\rangle dk.\label{emislontstaeq15}
\end{eqnarray}
Therefore, when the membrane is initially in a general density matrix
\begin{eqnarray}
\rho^{(b)}(0)=\sum_{m,n=0}^{\infty}\rho^{(b)}_{m,n}(0)|m\rangle_{b}\,_{b}\langle n|,\label{geninistaeq16}
\end{eqnarray}
the long-time state of the whole system can be obtained by superposition as
\begin{eqnarray}
\rho(\infty)=\sum_{m,n=0}^{\infty}\rho^{(b)}_{m,n}(0)\vert\varphi_{m}(\infty)\rangle\langle\varphi_{n}(\infty)|.\label{genscastaeq17}
\end{eqnarray}
Here the density matrix elements are $\rho^{(b)}_{m,n}(0)=\,_{b}\langle m|\rho^{(b)}(0)|n\rangle_{b}$.

To characterize the quadratic optomechanical coupling, a useful quantity
is the single-photon emission spectrum, i.e., the probability distribution of the emitted photon.
For the initial membrane state (\ref{geninistaeq16}), the emission spectrum is defined by
\begin{eqnarray}
S(\Delta_{k})&=&\textrm{Tr}[|1_{k}\rangle\langle 1_{k}|\rho(\infty)]\nonumber\\
&=&\sum_{l,m,n=0}^{\infty}\rho^{(b)}_{m,n}(0)B_{m,l,k}(\infty)B^{\ast}_{n,l,k}(\infty).\label{emisspecteq18}
\end{eqnarray}

\begin{figure}[tbp]
\center
\includegraphics[bb=17 14 374 394, width=3.3 in]{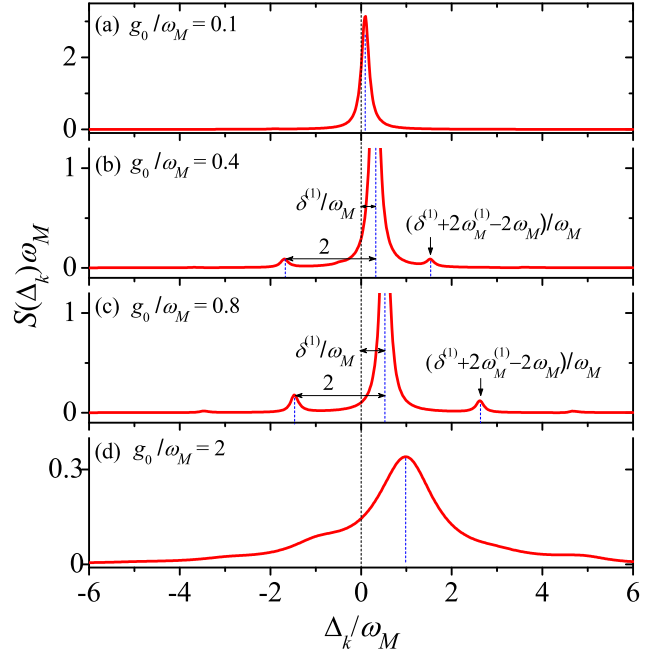}
\caption{(Color online) Single-photon emission spectrum $S(\Delta_{k})$ as a function of $\Delta_{k}$ for
various values of $g_{0}$ and $\gamma_{c}$. Panels (a-c) are plotted
in the resolved-sideband regime ($\gamma_{c}/\omega_{M}=0.2$), while (d) is plotted in the unresolved-sideband regime ($\gamma_{c}/\omega_{M}=1.5$). The membrane's
initial state is $|0\rangle_{b}$.} \label{emissspec_variousg}
\end{figure}
In Fig.~\ref{emissspec_variousg}, we plot the emission spectrum $S(\Delta_{k})$ versus the photon frequency $\Delta_{k}$, for
various values of $g_{0}$ and $\gamma_{c}$, when the membrane is initially in its ground state $|0\rangle_{b}$.
We see from Eq.~(\ref{emislontsoleq13}) that both $\omega^{(1)}_{M}>\gamma_{c}$ and $\omega_{M}>\gamma_{c}$ might be the
resolved-sideband condition. For a positive $g$, then $\omega_{M}>\gamma_{c}$ could make sure that the two
conditions are met, because of $\omega^{(1)}_{M}>\omega_{M}$. We found that, in the case of $\omega^{(1)}_{M}>\gamma_{c}>\omega_{M}$,
the phonon-sideband evidence is negligible. So, in this paper, we consider $\omega_{M}>\gamma_{c}$ as the resolved-sideband condition.
Figures~\ref{emissspec_variousg}(a-c)
are plotted in the resolved-sideband regime $\omega_{M}>\gamma_{c}$ so that the phonon sideband peaks could be used to characterize the coupling strength $g_{0}$.
When $g_{0}<\gamma_c$ [Fig.~\ref{emissspec_variousg}(a)], the spectrum is approximately a Lorentzian function
with width $\gamma_c$ and center $\Delta_{k}=\delta^{(1)}$. In this case, there are no sideband peaks in the spectrum.
However, the sideband peaks become visible when $g_{0}>\gamma_{c}$. Physically, when the displacement of the membrane
equals its zero-point fluctuation, the photon frequency shift induced by the quadratic optomechanical coupling is $g_{0}$.
To resolve this frequency shift from the Lorentzian spectrum of a free cavity, the condition $g_{0}>\gamma_{c}$
should be satisfied. Such a condition can also be understood by examining
the height of these peaks in the spectrum. To resolve a peak in the spectrum,
the peak height should be much higher than the tail of its neighboring Lorentzian. This requires $g_{0}\gg\gamma_{c}$
in the resolved-sideband regime. As an example, we analyze the special case of $g_{0}/\omega_{M}\ll 1$. In the
resolved-sideband regime $\omega_{M}/\gamma_{c}\gg 1$ and under the initial state $|0\rangle_{b}$, we expand
$S(\Delta_{k})$ up to second-order in $g_{0}/\omega_{M}$. Then, the height of the sideband peak located at
$\Delta_{k}=\delta^{(1)}-2\omega_{M}$ can be obtained as $S(\delta^{(1)}-2\omega_{M})\approx(\gamma_{c}/8\pi\omega^{2}_{M})(1+8g_{0}^{2}/\gamma^{2}_{c})$.
Since the main peak of the spectrum is approximately a Lorentzian function
$S_{L}(\Delta_{k})\approx\frac{\gamma_{c}}{2\pi}[(\Delta_{k}-\delta^{(1)})^{2}+\gamma^{2}_{c}/4]^{-1}$,
then the requirement
\begin{equation}
\frac{S(\delta^{(1)}-2\omega_{M})}{S_{L}(\delta^{(1)}-2\omega_{M})}\approx1+\frac{8g_{0}^{2}}{\gamma^{2}_{c}}\gg1\label{ratioeq19}
\end{equation}
leads to the condition $g_{0}\gg\gamma_{c}$.

\begin{figure}[tbp]
\center
\includegraphics[bb=17 14 374 394, width=3.3 in]{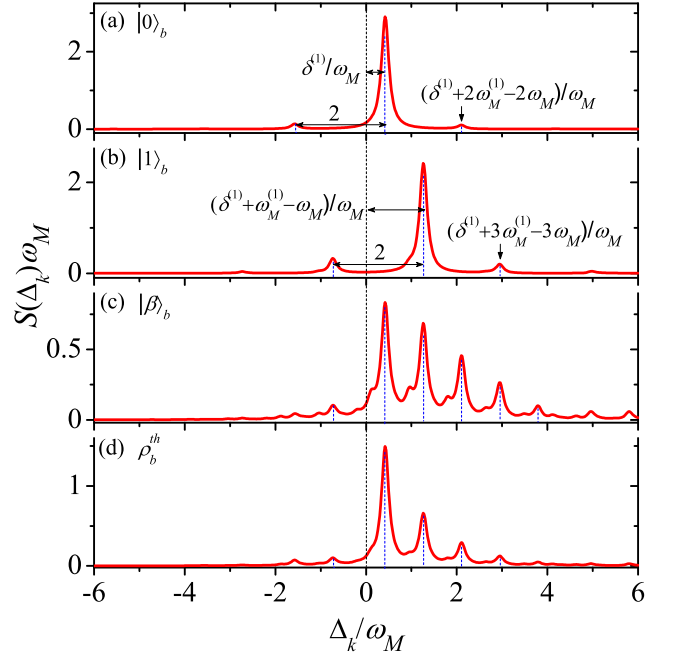}
\caption{(Color online) Single-photon emission spectrum $S(\Delta_{k})$ as a function of $\Delta_{k}$ when the
initial state of the membrane is either the Fock states $\vert 0\rangle_{b}$ and $\vert 1\rangle_{b}$, the coherent
state $\vert \beta=1\rangle_{b}$, or the thermal state $\rho^{th}_{b}(\bar{n}=1)$. Other parameters are
$\gamma_{c}/\omega_{M}=0.2$ and $g_{0}/\omega_{M}=0.6$.}\label{emissspec_initsta}
\end{figure}
We remark that the positions of these sideband peaks in Fig.~\ref{emissspec_variousg} are determined by the resonance condition~(\ref{emisrescondeq14}), and
these sideband peaks are not periodic because of the difference between $\omega_{M}^{(1)}$ and $\omega_{M}$.
Also, for the initial state $|0\rangle_{b}$ of the membrane, the contributing $m$ and $n$ in Eq.~(\ref{emisrescondeq14}) should be even numbers due to the parity requirement.
In Figs.~\ref{emissspec_variousg}(a-c), the peak located at $\Delta_{k}=\delta^{(1)}$ is the
main peak (corresponding to the transition $|1\rangle_{a}|\tilde{0}\rangle_{b}\rightarrow|0\rangle_{a}|0\rangle_{b}$).
Hence, we can resolve the main peak from the Lorentzian spectrum for a free cavity when $\delta^{(1)}>\gamma_{c}$, which requires
$g_{0}>\gamma_{c}(1+\gamma_{c}/\omega_{M})$. The peak located at $\Delta_{k}=\delta^{(1)}-2\omega_{M}$ corresponds to the
transition $|1\rangle_{a}|\tilde{0}\rangle_{b}\rightarrow|0\rangle_{a}|2\rangle_{b}$.
Moreover, the peak located at $\Delta_{k}=\delta^{(1)}+2\omega_{M}^{(1)}-2\omega_{M}$ corresponds to the transition
$|1\rangle_{a}|\tilde{2}\rangle_{b}\rightarrow|0\rangle_{a}|2\rangle_{b}$. Finally, Fig.~\ref{emissspec_variousg}(d) is plotted in the unresolved-sideband regime.
We can see from Fig.~\ref{emissspec_variousg}(d) that, even though the system works in the single-photon
strong-coupling regime, there are no sideband peaks in the spectrum. This fact indicates that
the resolved-sideband regime and the single-photon strong-coupling condition are two combined necessary requirements for observing
sideband peaks in the emission spectrum.

To illustrate how the spectrum depends on the initial state of the membrane,
we plot in Fig.~\ref{emissspec_initsta} the spectrum $S(\Delta_{k})$ versus $\Delta_{k}$ when the membrane is initially
in either the Fock states $\vert 0\rangle_{b}$ and $\vert 1\rangle_{b}$, the coherent state $\vert \beta=1\rangle_{b}$,
or the thermal state $\rho^{th}_{b}(\bar{n}=1)$. For the initial state $\vert 0\rangle_{b}$, the main peak (with the location $\Delta_{k}=\delta^{(1)}$) in Fig.~\ref{emissspec_initsta}(a) is related to the transition $|1\rangle_{a}|\tilde{0}\rangle_{b}\rightarrow|0\rangle_{a}|0\rangle_{b}$. The two peaks located at $\Delta_{k}=\delta^{(1)}-2\omega_{M}$ and $\Delta_{k}=\delta^{(1)}+2\omega^{(1)}_{M}-2\omega_{M}$ correspond to the transitions $|1\rangle_{a}|\tilde{0}\rangle_{b}\rightarrow|0\rangle_{a}|2\rangle_{b}$ and $|1\rangle_{a}|\tilde{2}\rangle_{b}\rightarrow|0\rangle_{a}|2\rangle_{b}$, respectively.
For the initial state $\vert 1\rangle_{b}$, the main peak (located at $\Delta_{k}=\delta^{(1)}+\omega^{(1)}_{M}-\omega_{M}$) in Fig.~\ref{emissspec_initsta}(b) is related to the transition $|1\rangle_{a}|\tilde{1}\rangle_{b}\rightarrow|0\rangle_{a}|1\rangle_{b}$. The other two peaks located at $\Delta_{k}=\delta^{(1)}+\omega^{(1)}_{M}-3\omega_{M}$ and $\Delta_{k}=\delta^{(1)}+3\omega^{(1)}_{M}-3\omega_{M}$ correspond to the transitions $|1\rangle_{a}|\tilde{1}\rangle_{b}\rightarrow|0\rangle_{a}|3\rangle_{b}$ and $|1\rangle_{a}|\tilde{3}\rangle_{b}\rightarrow|0\rangle_{a}|3\rangle_{b}$, respectively.
Here, Figs.~\ref{emissspec_initsta}(a) and (b) only show even- and odd-parity sideband peaks, respectively.
However, the coherent and thermal states contain both odd- and even-parity number states, and hence we can see both odd- and even-parity sideband peaks in Figs.~\ref{emissspec_initsta}(c,d). The positions of these sideband peaks are consistent with those in Figs.~\ref{emissspec_initsta}(a,b).

In the long-time limit, though the single photon is completely leaked out of the cavity, its state is still entangled with the mechanical mode.
This entanglement involves a single mode of phonons (the mechanical degree of freedom) and a set of modes of the photon
because the single photon is distributed into the continuous fields outside the cavity.
In general, it is difficult to clearly describe the structure of this entanglement. However, from the viewpoint of a single photon,
we can characterize the entanglement as a bipartite one between a single photon and a single mode of phonons. In particular,
we will consider a pure initial-state case so that the long-time state of the total system is also pure;
then we can employ the linear entropy to quantity this bipartite entanglement.

When the membrane is initially in the general state~(\ref{geninistaeq16}), the long-time state of the total system is given by Eq.~(\ref{genscastaeq17}).
In terms of Eq.~(\ref{emislontsoleq13}), the reduced density matrix of the membrane can be obtained as
\begin{eqnarray}
\rho^{(b)}(\infty)&=&i\gamma_{c}\sum_{l,l',m,n,s,s'=0}^{\infty}\rho^{b}_{m,n}(0)\Pi e^{-i(l-l')\omega_{M}t}\vert
l\rangle_{b}\,_{b}\langle l^{\prime}\vert,\label{lontstaofbmod20}
\end{eqnarray}
where
\begin{eqnarray}
\Pi=\frac{_{b}\!\langle l\vert\tilde{s}
\rangle_{b}\,_{b}\langle\tilde{s}\vert m\rangle_{b}\,_{b}\langle n\vert
\tilde{s'}\rangle_{b}\,_{b}\langle\tilde{s'}\vert l^{\prime}\rangle_{b}}{(l-l^{\prime})\omega_{M}+(s'-s)\omega_{M}^{(1)}+i\gamma_{c}}.\label{Pieq21}
\end{eqnarray}
The linear entropy~\cite{Zanardi2000} of the density matrix~(\ref{lontstaofbmod20}) is
\begin{eqnarray}
E_{l}&\equiv&1-\textrm{Tr}\{[\rho^{(b)}(\infty)]^{2}\}\nonumber\\
&=&1-\gamma _{c}^{2}\sum_{l,l^{\prime}=0}^{\infty}\left\vert
\sum_{m,n,s,s'=0}^{\infty}\rho^{b}_{m,n}(0)\Pi\right\vert^{2}.\label{linentemisseq22}
\end{eqnarray}

\begin{figure}[tbp]
\center
\includegraphics[bb=12 4 354 256, width=3.3 in]{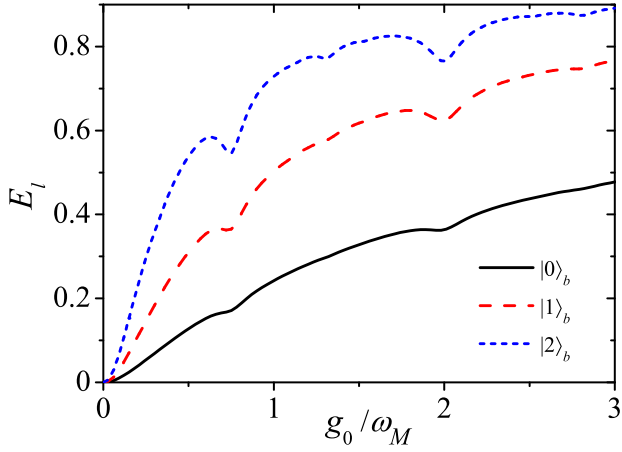}
\caption{(Color online) The linear entropy $E_{l}$ given in Eq.~(\ref{linentemisseq22}) versus the optomechanical coupling strength $g_{0}$, when the initial state of the membrane is $|0\rangle_{b}$, $|1\rangle_{b}$, and $|2\rangle_{b}$. Here $\gamma_{c}/\omega_{M}=0.2$.} \label{lineentropuemis}
\end{figure}
In Fig.~\ref{lineentropuemis}, we plot $E_{l}$ versus $g_{0}$ for initial Fock states $|n_{0}\rangle_{b}$ of the membrane, i.e., $\rho^{b}_{m,n}(0)=\delta_{m,n_{0}}\delta_{n,n_{0}}$.
Figure~\ref{lineentropuemis} shows that (as a general trend) $E_{l}$ increases with increasing $g_{0}$.
However, there are some resonance dips in the linear entropy when $g_{0}$ takes some special values.
The locations of these dips can be determined from the poles of the denominator in Eq.~(\ref{Pieq21}),
i.e., $(l-l^{\prime})\omega_{M}+(s'-s)\omega_{M}^{(1)}=0$, here the values of $(l-l^{\prime})$ and $(s'-s)$
should be even numbers because of the parity requirement in the transitions. For example, there is a dip
at $g_{0}/\omega_{M}=0.75$, which corresponds to $|(l-l^{\prime})/(s'-s)|=2$. In addition, corresponding to
the membrane's initial states $|0\rangle_{b}$, $|1\rangle_{b}$, and $|2\rangle_{b}$, the linear entropy keeps
increasing. We may roughly explain this phenomenon by analyzing the magnitude distribution of these factors
$_{b}\langle n_{0}|\tilde{s'}\rangle_{b}$ and $_{b}\langle \tilde{s}|n_{0}\rangle_{b}$. When $n_{0}$ changes
from $0$ to $2$, the magnitude distribution of these transitions elements $_{b}\langle n_{0}|\tilde{s'}\rangle_{b}$
(for different $s'$) becomes increasingly smoother. This implies that the number of contributing coefficients
increases, and hence the entanglement increases.

\section{Single-photon scattering}
In the single-photon scattering case, the single photon is initially in a
Lorentzian wave packet $\sqrt{\frac{\epsilon}{\pi}}\int_{0}^{\infty}\frac{1}{\Delta_{k}-\Delta_{0}+i\epsilon}dk$ in the outside fields, where $\Delta_{0}$
and $\epsilon$ are the detuning center and spectral width of the photon. In this case, with the Laplace transform method, the long-time solution (i.e., $t\gg 1/\gamma_{c},1/\epsilon$) of these probability amplitudes is obtained as $A_{n_{0},m}(\infty)=0$ and
\begin{eqnarray}
B_{n_{0},m,k}(\infty)&=&e^{-i(\Delta_{k}+m\omega_{M})t}\left(\sqrt{\frac{\epsilon}{\pi}}
\frac{1}{\Delta_{k}-\Delta_{0}+i\epsilon}\delta_{m,n_{0}}\right.\nonumber\\
&&\left.-\sqrt{\frac{\epsilon}{\pi}}
\frac{1}{\Delta_{k}-[\Delta_{0}+(n_{0}-m)\omega_{M}]+i\epsilon}\right.\nonumber\\
&&\left.\times\sum_{n=0}^{\infty}\frac{i\gamma_{c}\,_{b}\langle m\vert \tilde{n}
\rangle_{b}\,_{b}\langle\tilde{n}\vert n_{0}\rangle_{b}}{\Delta_{k}-\delta^{(1)}-n\omega_{M}^{(1)}+m\omega
_{M}+i\gamma_{c}/2}\right).\label{scaamplteq23}
\end{eqnarray}
Here the subscript $n_{0}$ in these amplitudes is used to mark the initial state of the membrane.
In the long-time limit, the single photon will completely leak out of the cavity, and hence we have $A_{n_{0},m}(\infty)=0$. It can be seen from $B_{n_{0},m,k}(\infty)$ that there are two physical processes in the single-photon scattering. (i) The single-photon direct-reflection process: the incident photon is directly reflected by the mirror, without entering the cavity. This process is described by the first term of $B_{n_{0},m,k}(\infty)$. (ii) The photon-membrane interacting process: the single photon enters the cavity to couple with the moving membrane and eventually leaks out of the cavity via the cavity decay channel. This process is described by the second term (i.e., the second and third line) in Eq.~(\ref{scaamplteq23}). In this process, the system experiences the transitions $|0\rangle_{a}|n_{0}\rangle_{b}\rightarrow|1\rangle_{a}|\tilde{n}\rangle_{b}\rightarrow|0\rangle_{a}|m\rangle_{b}$, These transitions are governed by the two resonance conditions
\begin{subequations}
\label{scarescondeq24}
\begin{align}
\Delta_{0}=&\delta^{(1)}+n\omega^{(1)}_{M}-n_{0}\omega_{M},\\
\Delta_{k}=&\delta^{(1)}+n\omega^{(1)}_{M}-m\omega_{M},
\end{align}
\end{subequations}
which can be derived from either the energy level structure in Fig.~\ref{setup}(b) or the poles of the probability amplitude~(\ref{scaamplteq23}). Interestingly, the second line in Eq.~(\ref{scaamplteq23}) is a Lorentzian wave packet with spectral width $\epsilon$ and center $\Delta_{k}=\Delta_{0}+(n_{0}-m)\omega_{M}$. In comparison to the initial Lorentzian wave packet, the shift of the wave packet center is equal to the energy variance of the membrane. Moreover, the third line in Eq.~(\ref{scaamplteq23}) has a similar form as Eq.~(\ref{emislontsoleq13}) for the single-photon emission process.
\begin{figure}[tbp]
\center
\includegraphics[bb=9 18 373 388, width=3.3 in]{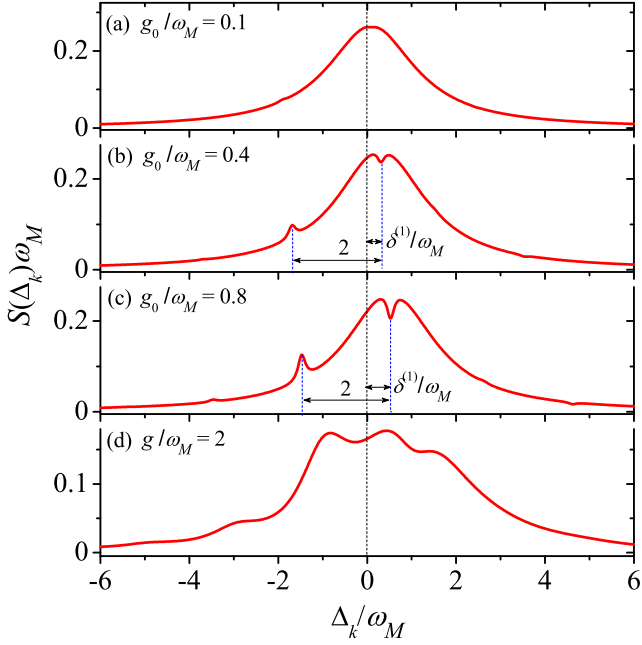}
\caption{(Color online) Single-photon scattering spectrum $S(\Delta_{k})$ versus the photon frequency $\Delta_{k}$ for various values of $g_{0}$ and $\gamma_{c}$. Figures (a-c) are plotted
in the resolved-sideband regime ($\gamma_{c}/\omega_{M}=0.2$), while (d) is plotted in the unresolved-sideband regime ($\gamma_{c}/\omega_{M}=1.5$). The membrane's initial state is $|0\rangle_{b}$. Other parameter are $\Delta_{0}=\delta^{(1)}$ and $\epsilon/\omega_{M}=1.2$.} \label{scattspec_variousg}
\end{figure}

The single-photon scattering spectrum can be calculated in terms of Eqs.~(\ref{emisspecteq18}) and~(\ref{scaamplteq23}). We see from Eq.~(\ref{scaamplteq23}) that either the second line or the third line could cause phonon sidebands, and
the conditions for resolving these sidebands due to the two lines are $\omega_{M}>\epsilon$ and $\omega_{M}>\gamma_{c}$, respectively. Here, $\gamma_{c}$ is the system's inherent parameter while $\epsilon$ is an externally controllable parameter. In the following, we first consider the case of $\epsilon>\omega_{M}>\gamma_{c}$ so that the observed sideband peaks are caused purely by the system inherent effect.
In Fig.~5, we plot the spectrum $S(\Delta_{k})$ versus the photon frequency $\Delta_{k}$ for various values of $g_{0}$ and $\gamma_{c}$.
When $g_{0}<\gamma_{c}$, there are no peaks in the spectrum [Fig.~\ref{scattspec_variousg}(a)]. The phonon sideband effect can be observed in the scattering spectrum when $g_{0}>\gamma_{c}$ and $\omega_{M}>\gamma_{c}$ [Figs.~\ref{scattspec_variousg}(b,c)]. Owing to the interference between the direct reflection process and the photon-membrane interacting process, there exist both peaks and dips in the spectrum. In Figs.~\ref{scattspec_variousg}(b,c), the dips represent the transition $|1\rangle_{a}|\tilde{0}\rangle_{b}\rightarrow|0\rangle_{a}|0\rangle_{b}$, while the peaks correspond to the transition $|1\rangle_{a}|\tilde{0}\rangle_{b}\rightarrow|0\rangle_{a}|2\rangle_{b}$.
In addition, in the unresolved-sideband regime ($\gamma_{c}>\omega_{M}$), there are no peaks even in the single-photon strong-coupling regime [Fig.~\ref{scattspec_variousg}(d)].
\begin{figure}[tbp]
\center
\includegraphics[bb=9 18 373 388, width=3.3 in]{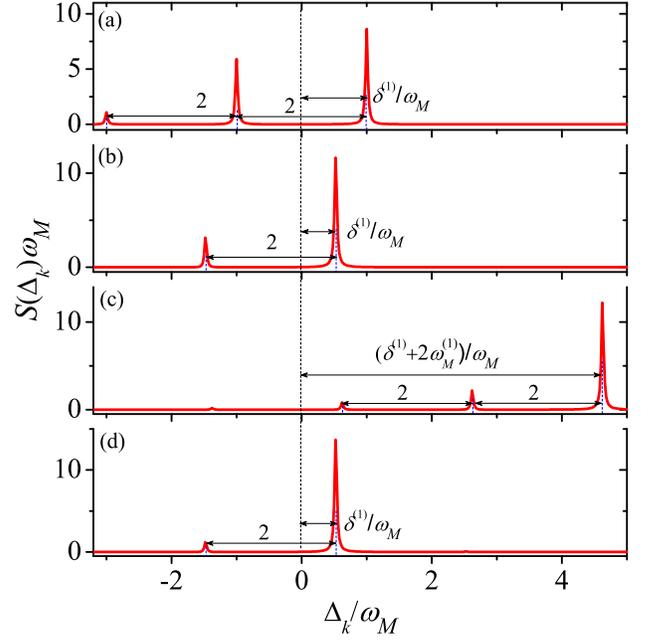}
\caption{(Color online) Scattering spectrum $S(\Delta_{k})$ versus $\Delta_{k}$. The panel (a) is plotted in the unresolved-sideband regime ($g_{0}/\omega_{M}=2$ and $\gamma_{c}/\omega_{M}=1.5$), while panels (b-d) are plotted in the resolved-sideband regime ($g_{0}/\omega_{M}=0.8$ and $\gamma_{c}/\omega_{M}=0.2$). The frequency center $\Delta_{0}$ of the incident photon is $\delta^{(1)}+2\omega^{(1)}_{M}$ in (c), and $\delta^{(1)}$ in other panels. The membrane's initial state is the ground state $|0\rangle_{b}$ in panels (a-c) and the coherent state $|\alpha=1\rangle_{b}$ in (d). The parameter $\epsilon/\omega_{M}=0.02$.} \label{scatspectrum2}
\end{figure}

We now consider the near-monochromatic case ($\epsilon\ll\gamma_{c}$). In Fig.~\ref{scatspectrum2}(a), we plot the scattering spectrum in the case of $\gamma_{c}>\omega_{M}$ and $\epsilon\ll\omega_{M}$. This figure exhibits phonon sideband peaks, and hence indicates that $\omega_{M}>\epsilon$ also provides the condition for observing the phonon sideband peaks due to the second line in Eq.~(\ref{scaamplteq23}).
We point out that this provides a way to characterize the coupling strength $g_{0}$ from the scattering spectrum in the case of $\gamma_{c}>\omega_{M}$.
Another benefit in the near-monochromatic case is that we can conveniently control the exciting transition by choosing the frequency of the incident photon. In Figs.~\ref{scatspectrum2}(b,c), we choose the frequency of the incident photon as $\Delta_{0}=\delta^{(1)}$ and $\Delta_{0}=\delta^{(1)}+2\omega_{M}^{(1)}$ to resonantly excite the system from $|0\rangle_{a}|0\rangle_{b}$ to $|1\rangle_{a}|\tilde{0}\rangle_{b}$ and $|1\rangle_{a}|\tilde{2}\rangle_{b}$, respectively. In the emission process, the membrane will experience the transitions from $|\tilde{0}\rangle_{b}$ and $|\tilde{2}\rangle_{b}$ to $|n\rangle_{b}$ $(n=0,2,4,...)$. Therefore, the maximal frequency sideband peaks should be located at $\Delta_{k}=\delta^{(1)}$ and $\delta^{(1)}+2\omega_{M}^{(1)}$, respectively. In addition, the period of these peaks is $2\omega_{M}$. Similar to the emission case, the scattering spectrum also depends on the initial state of the membrane. In Fig.~\ref{scatspectrum2}(d), we plot the scattering spectrum when the membrane's initial state is coherent state $|\alpha=1\rangle_{b}$. Though the initial coherent state contains both even- and odd-parity states, the spectrum only exhibits similar peaks as those in Fig.~\ref{scatspectrum2}(b). This is because the incident photon (with $\Delta_{0}=\delta^{(1)}$) only resonantly excites the membrane from $|0\rangle_{b}$ to $|\tilde{0}\rangle_{b}$; other transitions from $|n\rangle_{b}$ to $|\tilde{n}'\rangle_{b}$ ($n,n'=1,2,3,...$, with the same parity) are significantly suppressed due to the large detuning. A further photon emission process induces the transitions from $|\tilde{0}\rangle_{b}$ to $|n\rangle_{b}$ $(n=0,2,4,...)$.
\begin{figure}[tbp]
\center
\includegraphics[bb=6 34 416 536, width=3.3 in]{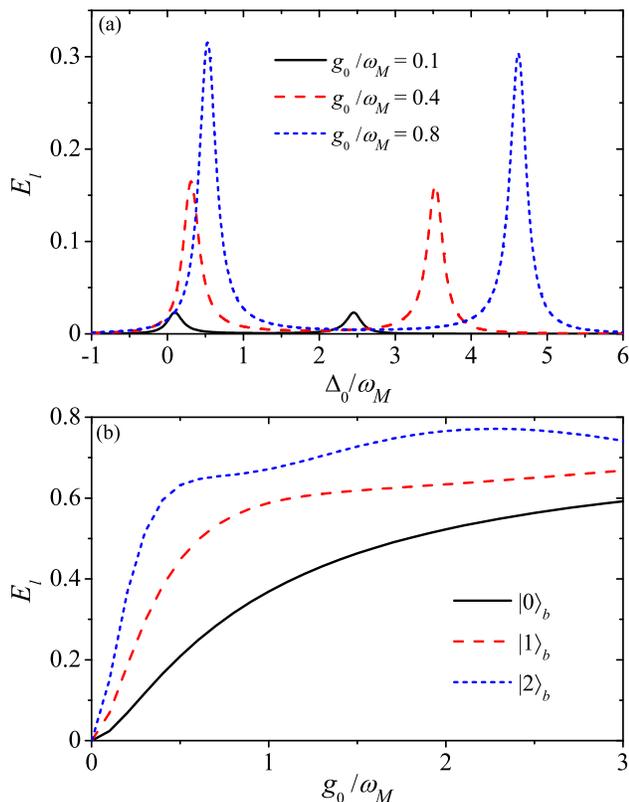}
\caption{(Color online) (a) The linear entropy $E_{l}$ versus $\Delta_{0}$ for various values of $g_{0}$. The initial state of the membrane is $|0\rangle_{b}$.
(b) The linear entropy $E_{l}$ versus $g_{0}$ when the membrane is initially in states $|0\rangle_{b}$, $|1\rangle_{b}$, and $|2\rangle_{b}$. Here the frequencies of the incident photon are $\Delta_{0}=\delta^{(1)}$, $\Delta_{0}=\delta^{(1)}+\omega_{M}^{(1)}-\omega_{M}$, and $\Delta_{0}=\delta^{(1)}+2\omega_{M}^{(1)}-2\omega_{M}$, respectively. Other parameters are $\gamma_{c}/\omega_{M}=0.2$ and $\epsilon/\omega_{M}=0.02$.} \label{lineentropscattering}
\end{figure}

Similar to the emission case, the scattered photon is completely emitted out of the cavity in the long-time limit;
and the state of the photon is entangled with the mechanical membrane. For the membrane's initial state $|n_{0}\rangle_{b}$,
the linear entropy of the long-time reduced density matrix of the membrane was calculated exactly.
This is not shown here because the analytical solution is long. We can examine how the linear entropy depends on the system
parameters: the incident photon frequency $\Delta_{0}$ and the optomechanical coupling strength $g_{0}$.
In the near-monochromatic limit $\epsilon\ll\gamma_{c}$, we plot, in Fig.~\ref{lineentropscattering}(a), the linear entropy $E_{l}$ as a function of $\Delta_{0}$
when the quadratic optomechanical coupling $g_{0}$ takes various values.
We can see from Fig.~\ref{lineentropscattering}(a) that there are resonant peaks in the entropy. For a given $g_{0}$, the locations of these peaks are $\Delta_{0}=\delta^{(1)}$ and $\Delta_{0}=\delta^{(1)}+2\omega_{M}^{(1)}$, which are determined by the resonance conditions in the dominant transitions in the photon injection process: $|0\rangle_{a}|0\rangle_{b}\rightarrow|1\rangle_{a}|\tilde{0}\rangle_{b}$ and $|0\rangle_{a}|0\rangle_{b}\rightarrow|1\rangle_{a}|\tilde{2}\rangle_{b}$.

We also investigate the dependence of the linear entropy $E_{l}$ on the coupling strength $g_{0}$ when the incident photon is in resonance with the transitions. In Fig.~\ref{lineentropscattering}(b), we plot the entropy $E_{l}$ versus $g_{0}$ when the initial state of the membrane is $|0\rangle_{b}$, $|1\rangle_{b}$, and $|2\rangle_{b}$. Here the single photon is resonantly injected into the cavity. Corresponding to the initial states $|0\rangle_{b}$, $|1\rangle_{b}$, and $|2\rangle_{b}$, the driving frequencies are $\Delta_{0}=\delta^{(1)}$, $\Delta_{0}=\delta^{(1)}+\omega_{M}^{(1)}-\omega_{M}$, and $\Delta_{0}=\delta^{(1)}+2\omega^{(1)}_{M}-2\omega_{M}$, respectively. These drivings determine the dominant photon-injection transitions: $|0\rangle_{a}|0\rangle_{b}\rightarrow|1\rangle_{a}|\tilde{0}\rangle_{b}$, $|0\rangle_{a}|1\rangle_{b}\rightarrow|1\rangle_{a}|\tilde{1}\rangle_{b}$, and $|0\rangle_{a}|2\rangle_{b}\rightarrow|1\rangle_{a}|\tilde{2}\rangle_{b}$. We can see that, in the resonant scattering case, the linear entropy increases when increasing $g_{0}$.

\section{Discussions and conclusions}

Although the currently-available quadratic couplings are too weak to reach the single-photon
strong-coupling regime, advances have recently been made in the enhancement of this coupling strength. In quadratic optomechanics, the coupling strength is
$g_{0}=\eta x^{2}_{\rm{zpf}}$, where $x_{\rm{zpf}}$ is the zero-point fluctuation of the mechanical
membrane, and $\eta=\frac{1}{2}\frac{\partial^{2}\omega_{c}(x)}{\partial x^{2}}|_{x=0}$, with $\omega_{c}(x)$ being the $x$-dependent cavity frequency.
Recently, the value of $\eta$ has been increased significantly from about $30$ MHz/nm$^2$ (in Ref.~\cite{Harris2010})
to $20$ GHz/nm$^2$ (in Ref.~\cite{Flowers-Jacobs2012}) using a fiber cavity with a smaller mode size.
For a $x_{\rm{zpf}}\sim 41$ fm, suggested in Ref.~\cite{Harris2008}, the coupling strength is $g_{0}\sim2\pi\times5.35$ Hz.
If the cavity decay rate $\gamma_{c}\sim$ MHz, the coupling strength needs to be further increased by five orders of
magnitude to reach the single-photon strong-coupling regime. From $g_{0}=\eta x^{2}_{\rm{zpf}}$, we can see that
the $g_{0}$ can be increased by obtaining a larger $\eta$ or $x_{\rm{zpf}}$.
This requires the improvement of experimental conditions.
In addition, the model under investigation is a general quadratic optomechanical Hamiltonian\cite{Vanner2011}.
It can also be realized in various physical systems such as ultracold atoms and superconducting circuits.
In the linear optomechanical coupling case, the ultracold atom system has been demonstrated to approach
the single-photon strong-coupling regime\cite{Murch2008,Brennecke2008}, and the
superconducting circuit system has been estimated to be in this regime\cite{Sillanpaa2013,Rimberg2014}.
Moreover, other methods have been recently explored to achieve effective strong quadratic optomechanics. For example,
a measurement-based method has been proposed to perform this mission.
Based on these achievements,
it might be possible to pursue the single-photon strong-coupling regime in quadratic optomechanics.

In the above discussions, we did not include the mechanical dissipation. Now we give a rough estimate for the influence of the mechanical dissipation on our results.
In our considerations, all the results are determined by the probability amplitudes given in Eqs. (\ref{emislontsoleq13}) and (\ref{scaamplteq23}).
To evaluate the influence of the mechanical dissipation, we introduce an imaginary decay factor into the resonant frequency
of the mechanical mode, i.e., approximately replacing $\omega_{M}$ with $\omega_{M}-i\gamma_{M}(2\bar{n}_{th}+1)/2$,
where $\bar{n}_{th}$ is the thermal phonon occupation. In this way we can estimate the effect of the thermal mechanical
dissipation. If $\gamma_{M}(2\bar{n}_{th}+1)\ll\gamma_{c}/2$, then the mechanical dissipation is negligible. This is because, in the low-temperature regime,
the imaginary part in the denominators of Eqs. (\ref{emislontsoleq13}) and (\ref{scaamplteq23}) can still be approximated by $\gamma_{c}/2$, and the
exponential factor $e^{-\gamma_{M}(2\bar{n}_{th}+1)t}$ is almost $1$ during the time scale $t\sim 1/\gamma_{c}$.

We have analytically studied the single-photon emission and scattering
in a quadratically-coupled optomechanical system. By treating the optomechanical cavity and
its environment as a whole system, we have obtained the emission and scattering
solutions using the Laplace transform method. Based on our solutions, we have calculated the single-photon
emission and scattering spectra, and found relations between the spectral features and the system's inherent
parameters. In particular, we have clarified the condition under which phonon sideband peaks can be observed in the photon spectra. In the resolved-sideband regime $\omega_{M}>\gamma_{c}$,
the phonon sidebands are visible when $g_{0}>\gamma_{c}$, while the
condition for resolving the photon-state energy-level shift $\delta^{(1)}$ is $g_{0}>\gamma_{c}(1+\gamma_{c}/\omega_{M})$.
We have also investigated the creation of photon-phonon entanglement in the emission and scattering processes.
This entanglement was created due to the energy requirement in the photon absorption and emission processes,
and it was treated as a bipartite one from the viewpoint of a single photon rather than photon modes.
We have considered the pure state case so that the linear entropy can be used to characterize this entanglement between the phonon mode and a single photon.

Finally, we want to mention several possible applications of the single-photon emission and scattering processes. (i) Since the emission and scattering spectra are related to the
eigen-energy of the system, we can read out the system parameters, such as coupling strength and mechanical frequency, from the spectra. (ii) We can infer the initial state information of the
mechanical resonator based on the single-photon spectra, similar to the linear optomechanics case\cite{Liao2014}. As a special example, the initial temperature of the mechanical membrane can also be determined from the spectra. (iii) The entanglement generated in the emission and scattering processes involves phonons and one outgoing photon. So this entanglement could be used to implement quantum information processing tasks, where phonons and photons are used as information memory and carriers, respectively.

\begin{acknowledgments}
J.Q.L. is supported by the Japan Society for the Promotion of Science
(JSPS) Foreign Postdoctoral Fellowship No. P12503. F.N. is partially supported by the RIKEN iTHES Project,
MURI Center for Dynamic Magneto-Optics, and a Grant-in-Aid for Scientific Research (S).
\end{acknowledgments}

\end{document}